\documentclass[aps,showpacs]{revtex4-1}
\usepackage{bm}
\usepackage{epsfig}
\usepackage{amsmath}

\def\be{\begin{equation}}
\def\lan{\left\langle}
\def\ran{\right\rangle}
\def\ee{\end{equation}}
\def\barr{\begin{array}}
\def\earr{\end{array}}

\def\nn8{\\}
\def\l{\left}
\def\r{\right}
\def\dis{\displaystyle}
\def\ed{\end{document}}

\def\w{\omega}
\def\dis{\displaystyle}
\def\e{\varepsilon}

\begin{document}

\title{Poisson to GOE transition in the distribution of the ratio of consecutive
level spacings}

\author{N. D. Chavda$^1$, H. N. Deota$^1$ and V. K. B. Kota$^2$}
\affiliation{$^1$Applied Physics Department, Faculty of Technology and
Engineering,\\ Maharaja Sayajirao University of Baroda, Vadodara 390 001, India\\
$^2$Physical Research Laboratory, Ahmedabad 380 009, India}

\begin{abstract}

Probability distribution for the  ratio ($r$) of consecutive level spacings of
the eigenvalues of a Poisson (generating  regular spectra) spectrum and that of
a GOE random matrix ensemble are given recently. Going beyond these, for the
ensemble generated by the  Hamiltonian $H_\lambda = (H_0+\lambda
V)/\sqrt{1+\lambda^2}$ interpolating Poisson ($\lambda=0$) and GOE ($\lambda
\rightarrow \infty$) we have analyzed the transition curves for $\lan r\ran$ and $\lan
\tilde{r}\ran$ as $\lambda$ changes from $0$ to $\infty$; $\tilde{r} =
min(r,1/r)$.  Here, $V$ is a GOE ensemble of real symmetric $d \times d$
matrices and $H_0$ is a diagonal matrix with a Gaussian distribution (with mean
equal to zero) for the diagonal matrix elements; spectral variance generated by
$H_0$ is assumed to be same as the one generated by $V$. Varying $d$ from 300 to
1000, it is shown  that the transition parameter is  $\Lambda \sim
\lambda^2\,d$, i.e. the $\lan r\ran$ vs  $\lambda$ (similarly for $\lan \tilde{r}\ran$ vs $\lambda$) curves for  different $d$'s merge to
a single curve when this is considered as a function  of $\Lambda$.
Numerically, it is also found that this transition curve  generates a mapping to a
$3 \times 3$ Poisson to GOE random matrix ensemble. Example for Poisson to GOE
transition from a one dimensional interacting spin-1/2 chain is presented.

\end{abstract}

\date{\today}
\maketitle

\section{Introduction}
\label{sec:1}

Energy level fluctuations in the quantum systems whose classical analogue is chaotic
will follow one of the three classical ensembles, the Gaussian orthogonal ensemble
(GOE), Gaussian unitary ensemble(GUE) or Gaussian symplectic ensemble(GSE) depending on the symmetries of
the Hamiltonian~\cite{Haake,stockmann,wimberger}. The nearest neighbor spacing distribution (NNSD)
$P(S)dS$ giving the degree of level repulsion is one of the commonly used measures
for level statistics. More importantly, Berry and Tabor established that if a quantum
system is in the integrable domain (corresponding to regular behavior), the NNSD follows
the Poisson distribution [$P(S) =\exp ( -S)$]. However, as conjectured by Bohigas et al \cite{Bohigas}, if the system
(where the Hamiltonian preserves time-reversal and rotational invariance) is chaotic then
the NNSD will be described by the Wigner surmise which is essentially the GOE result
[$P(S) = (\pi/2) S \exp (-\pi S^2/4)$].

Recently, the $P(r)$ distribution of the ratio of consecutive level spacings ($r$) and the related averages
(for example $\lan r\ran$) are established to be useful statistical measures to distinguish `order' and `chaos'
in the energy levels. It is so, because here no unfolding of energy spectrum is required as they are independent
of the form of the density of energy levels. The $P(r)$ and $\lan r \ran$ will allow for
a more transparent comparison with  experimental results than the traditional
NNSD. These are used to investigate  numerically many-body localization in
mesoscopic systems \cite{Huse2007,OPH2009,Pal-10,Iyer-12}, to quantify the
distance from integrability on finite size lattices appropriate for ultra cold
atomic gases \cite{koll2010,Coll2012}, and also to establish, using embedded
random matrix ensembles, that finite many particle quantum systems with strong enough
interactions follow GOE \cite{CK}. More recently, they are used in the study of
spectral  correlations in diffuse van der Waals clusters \cite{HBCK} and in the analysis of cancer networks \cite{RMJ}.

Most of the many-particle quantum systems such as atoms, nuclei, quantum dots,
small metallic grains and so on are mixed systems with level fluctuations
intermediate to Poisson and GOE (in this paper we will restrict to systems with
time reversal and rotational symmetry). In order to quantify the degree of chaos
in a complex system, random matrix ensembles interpolating between Poisson and GOE have
been considered (see for example \cite{Haake,rmt2,rmt3,vkbk01}) and  the formulas
for NNSD are derived. One of the most popular formula is the Brody distribution
\cite{Br-72} and other that is used in deciding the Poisson to GOE  transition
marker is derived using a $2 \times 2$ interpolating matrix ensemble
\cite{Ko-14}. The NNSD bears the process of unfolding the spectrum which isn't the case in $P(r)$ and hence the $P(r)$ is a
more practical measure of level fluctuations. Therefore, it is important to study Poisson to GOE
transition in $P(r)$ for a random matrix ensemble that interpolates the two
limiting distributions. A simple interpolation formula was
suggested in \cite{CK} for Poison to GOE transition in $P(r)$ by drawing an analogy with the Brody distribution. However, it is possible to
write many more interpolating formulas. Because of this, it is important to derive
$P(r)$ and related averages using a random matrix ensemble which interpolates between the two limits. The main purpose of this letter is to study the transition from Poisson to GOE in terms of the $P(r)$ distribution and the related averages. Now we will give a preview.

In Section~\ref{sec:2}, to get started, the analytical results for the Poisson and the GOE for the
probability distribution $P(r)$ of the ratio of consecutive level spacings and
the related averages $\lan r\ran$ and $\lan \tilde{r} \ran$ ($\tilde{r}$ is defined
in Section~\ref{sec:2}) are briefly discussed. Then, a $3 \times 3$ interpolating
random matrix ensemble {$H_{3 \times 3}$} is introduced which gives Poisson and GOE at
the limits. Note that, the minimal size of the matrix required for the study of $P(r)$ is $3 \times 3$.
In Section~\ref{sec:3}, a more general interpolating ensemble $\{H_\lambda\}$
for $d \times d$ matrices is introduced and using 1000 member ensembles with $d
\leq 1000$, numerically studied are $\lan r\ran$ and $\lan \tilde{r} \ran$ as a
function of the $\lambda$ parameter. Established here are (i) the transition
parameter $\Lambda \sim \lambda^2 d$ giving scaling, (ii) the universal form for
the transition curves $\lan r\ran$,  $\lan \tilde{r} \ran$ vs $\Lambda$ and (iii)
 mapping between $d \times d$ and $3 \times 3$ matrix ensembles.  Section~\ref{sec:4} gives
 a simple application of the transition curves and the results for Poisson to GOE
transition in a 1D interacting spin-1/2 chain. Finally, Section~\ref{sec:5}
gives conclusions and future outlook.

\section{Poisson and GOE formulas for $P(r)$ and a $3 \times 3$ ensemble for
Poisson to GOE transition}
\label{sec:2}

Let us consider an ordered set of eigenvalues (energy levels) $E_n$, where
$n=1,2,...,d$. The nearest-neighbor spacing is given by $s_n = E_{n+1} - E_{n}$.
Then, the ratio of two consecutive level spacings is $r_n=s_{n+1}/s_n$. The
probability distribution for consecutive level spacings is denoted by $P(r)dr$.
If the system is in the integrable domain, the nearest-neighbor spacings follow Poisson and  then
$P(r)$ is \cite{Huse2007,ABGR-2013} [denoted by $P_P(r)$],
\be
P_P(r)=\dis\frac{1}{(1+r)^2}\;.
\label{eq1}
\ee
Similarly, the $P(r)$ derived using $3 \times 3$ real symmetric matrices for GOE is given by a Wigner-like surmise
\cite{ABGR-2013},
\be
P_W(r)= \frac{27}{8} \frac{r+r^2}{(1+r+r^2)^{5/2}} \;.
\label{eq2}
\ee
In addition to $r_n$, it is possible to consider the distribution of the ratios
$\tilde{r}_n$ where $\tilde{r}_n = \frac{\min(s_n,s_{n-1})}{\max(s_n,s_{n-1})} =
\min(r_n,1/r_n)$. As pointed out in \cite{ABGR-2013}, it is possible to write
down  $P(\tilde{r})$ for given $P(r)$. In practice, besides $P(r)$ it is also useful to consider
$\lan r\ran$ and $\lan \tilde{r}\ran$  corresponding to the average value of $r$ and $\tilde{r}$ respectively. For GOE,
$\lan r \ran= 1.75$ and for Poisson it is $\infty$. Similarly, $\lan \tilde{r} \ran = 0.536$ for GOE and
$0.386$ for  Poisson. We will use $P(r)$, $\lan r \ran$ and $\lan \tilde{r}\ran$
in the following discussion. Before considering Poisson to GOE transition, let us add that
going beyond GOE, formulas for $P(r)$ and also for the probability
distribution of more generalized ratios of spacings have been derived recently \cite{Atas13}
for $\beta$-Hermite and $\beta$-Laguerre ensembles.

For the Poisson to GOE transition in NNSD, a simple $2 \times 2$ ensemble was
constructed in \cite{Ha-91,Karol-91,KS-99} and it was shown that the results of
this ensemble can be mapped to those of any $d \times d$ matrix. Following this,
it is possible to identify a $3 \times 3$ ensemble for Poisson to GOE
interpolation in $P(r)$ and the related averages. For this, we will consider the
following $3 \times 3$ matrix ensemble $H_{3\times3}$,
\be
H_{3 \times 3} = \l[\barr{ccc} A & B & C \\ B & D & E \\
C & E & F \earr\r] = \l[\barr{ccc} -pvx\;\;&0\;\;&0 \\ 0\;\;&0\;\;&0 \\ 0\;\;&
0\;\;& pvy \earr \r] \;+\;\;\lambda\;\l[\barr{ccc} a & b & c \\ b & d & e \\
c & e & f \earr\r] \;.
\label{eq3}
\ee
In Eq.~(\ref{eq3}), $x$ and $y$ are independent Poisson variables with average
unity so that the joint probability distribution $P(x,y)\,dxdy =
\exp-(x+y)\,dxdy$ with $0 \leq x,y \leq \infty$. Similarly, $a$, $b$, $c$, $d$,
$e$ and $f$ are independent Gaussian variables with zero center and variance equal to
$2v^2$ for $a$, $d$ and $f$ [i.e. they are independent G(0,$2v^2)$ variables]
while it is $v^2$ for $b$, $c$ and $e$ [i.e. they are independent G(0,$v^2$)
variables]. Also, in Eq.~(\ref{eq3}) $\lambda$ is a parameter and $p$ is a
constant introduced following the Poisson to GOE transition in NNSD analyzed using a $2\times 2$ ensemble [22]. Presence
of the $p$ will help in identifying the role of the mean spacing of the unperturbed spectrum (See Eq.(5) ahead). For
$\lambda=0$, $H_{3 \times3}$ gives the Poisson result for $P(r)$ as shown in Eq.~(\ref{eq1}) . Similarly, in the
limit $\lambda \rightarrow \infty$, $H_{3 \times3}$ gives the GOE result for $P(r)$ as shown in Eq.~(\ref{eq2}). Thus,
as $\lambda$ changes from $0$ to $\infty$, $H_{3\times 3}$ generates Poisson to GOE transition in $P(r)$ and the related
averages. Firstly, it is easy to see that in the limit $\lambda=0$, the average
spacing between the nearest eigenvalues is $\overline{D_0}=pv$. Now, the joint probability distribution
, $\rho(A,B,C,D,E,F)$, for the matrix elements $A$, $B$, $C$, $D$, $E$ and $F$ of $H_{3 \times 3}$ is easy to write down. Let the
eigenvalues of $H_{3 \times3}$ are $e_i$ ($i=1,2$ and $3$) and the orthogonal matrix that diagonalizes $H_{3 \times3}$  is
generated by three angles and say they are $\theta_i$ ($i=1,2$ and $3$). Then,
the joint probability distribution is given by
$$
\rho(A,B,C,D,E,F)={\cal N} \exp-\l\{\dis\frac{(A+pvx)^2 +D^2 + (F-pvy)^2}{
4\lambda^2v^2}+\dis\frac{B^2+C^2+E^2}{\lambda^2v^2}\r\}
$$
and $dAdBdCdDdEdF$ is of the form $\Pi_{i <j} \l|e_i - e_j\r|\;f(\theta_1,
\theta_2,\theta_3) de_1 de_2 de_3 d\theta_1 d\theta_2 d\theta_3$. Note that,
${\cal N}$ is a normalization constant and the function $f(\theta_1, \theta_2, \theta_3)$ follows from the
Jacobi transformation from $(A,B,C,D,E,F)$ to $(e_1, e_2, e_3, \theta_1,
\theta_2, \theta_3)$. Putting $e_i=2\lambda v\,x_i$, we have
\be
\barr{l}
\rho(x_1, x_2, x_3, \theta_1, \theta_2, \theta_3)\; dx_1 dx_2 dx_3 d\theta_1
d\theta_2 d\theta_3 = \\
{\cal N}^\prime \exp-\l\{\dis\sum_{i=1}^{3} x_i^2 +
\dis\frac{p^2v^2}{4\lambda^2v^2} (x^2 + y^2) + \dis\frac{pv}{\lambda v}\l[x
f_1(x_1, x_2, x_3, \theta_1, \theta_2, \theta_3) -y g_1(x_1, x_2, x_3,
\theta_1, \theta_2, \theta_3)\r]\r\} \\
\times\;\dis\Pi_{i <j} \l|x_i - x_j\r|\;f(\theta_1,
\theta_2, \theta_3) dx_1 dx_2 dx_3 d\theta_1 d\theta_2 d\theta_3 \;.
\earr \label{eq4}
\ee
Note that, $f_1$ and $g_1$ in Eq.~(\ref{eq4}) transform $A$ and $F$ in Eq.~(\ref{eq3})
 into $x_i$ (i.e. $e_i$) and $\theta_i$, whereas ${\cal N}^\prime$ is a
normalization constant. Integrating the R.H.S. of Eq.~(\ref{eq4}) over
$(\theta_1, \theta_2, \theta_3)$ and also over $x$ and $y$ with the weight factor $P(x,y)=\exp-(x+y)$ will
give $\rho(x_1, x_2, x_3)$ and there by $P(r)$. This problem we could not solve
so far. However, Eq.~(\ref{eq4}) gives the important result that   $\rho(e_1,
e_2, e_3)$ and therefore the $P(r)$ will depend only the transition parameter
$\Lambda$ where
\be
\Lambda = \dis\frac{\lambda^2 v^2}{p^2 v^2} = \dis\frac{\lambda^2
v^2}{\l[\;\overline{D_0}\;\r]^2} \;.
\label{eq5}
\ee
This is nothing but the square of the admixing matrix element divided by the
average spacing between the unperturbed ($\lambda=0$) levels. Now we will
consider a general $d \times d$ matrix for Poisson to GOE transition and show
that this ensemble can be mapped to the $H_{3 \times 3}$ ensemble.

\begin{figure}[tbh]
\begin{center}
\includegraphics[width=0.4\linewidth]{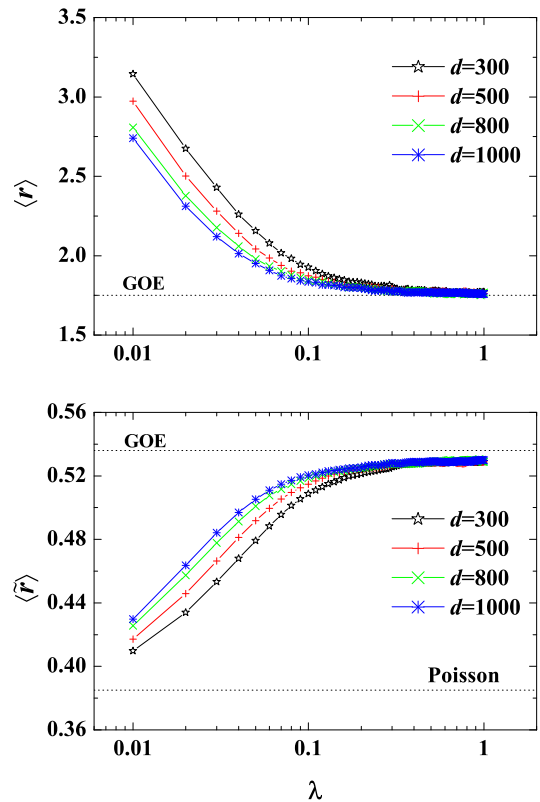}
\end{center}

\caption{(Color online) Variation of $\lan r \ran$ (upper panel) and $\lan \tilde{r} \ran$
(lower panel) as a function of $\lambda$ for $d\times d$ matrix ensembles
defined by $H_\lambda$ in Eq.~(\ref{eq6}). Results are shown for matrix dimension $d$ going from $300$ to $1000$. In each calculation, en ensemble of
1000 members is used.}
\label{fig1}
\end{figure}

\begin{figure}[tbh]
\begin{center}
\includegraphics[width=0.4\linewidth]{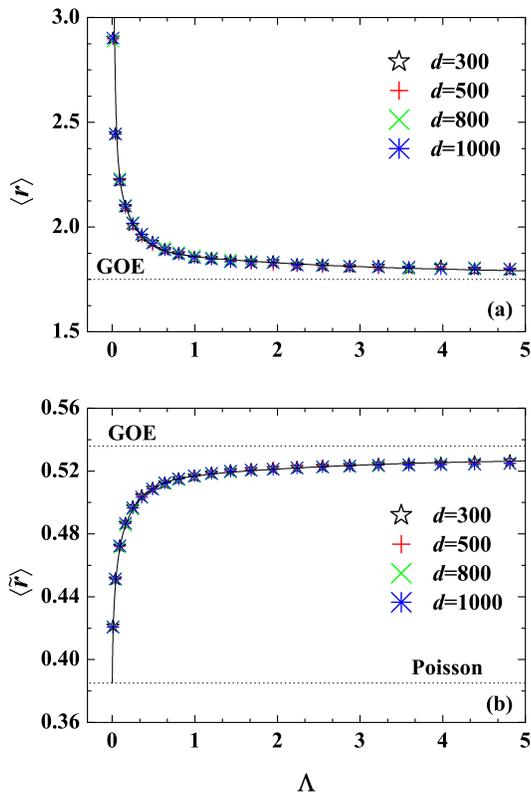}
\end{center}

\caption{(Color online) Same as Fig.~\ref{fig1} but for $\lan r \ran$ (upper panel) and $\lan \tilde{r}
\ran$ (lower panel) vs $\Lambda$. The continuous curves represent the best fit to the results giving
the universal transition curves.}
\label{fig2}
\end{figure}

\begin{figure}[tbh]
\begin{center}
\includegraphics[width=0.4\linewidth]{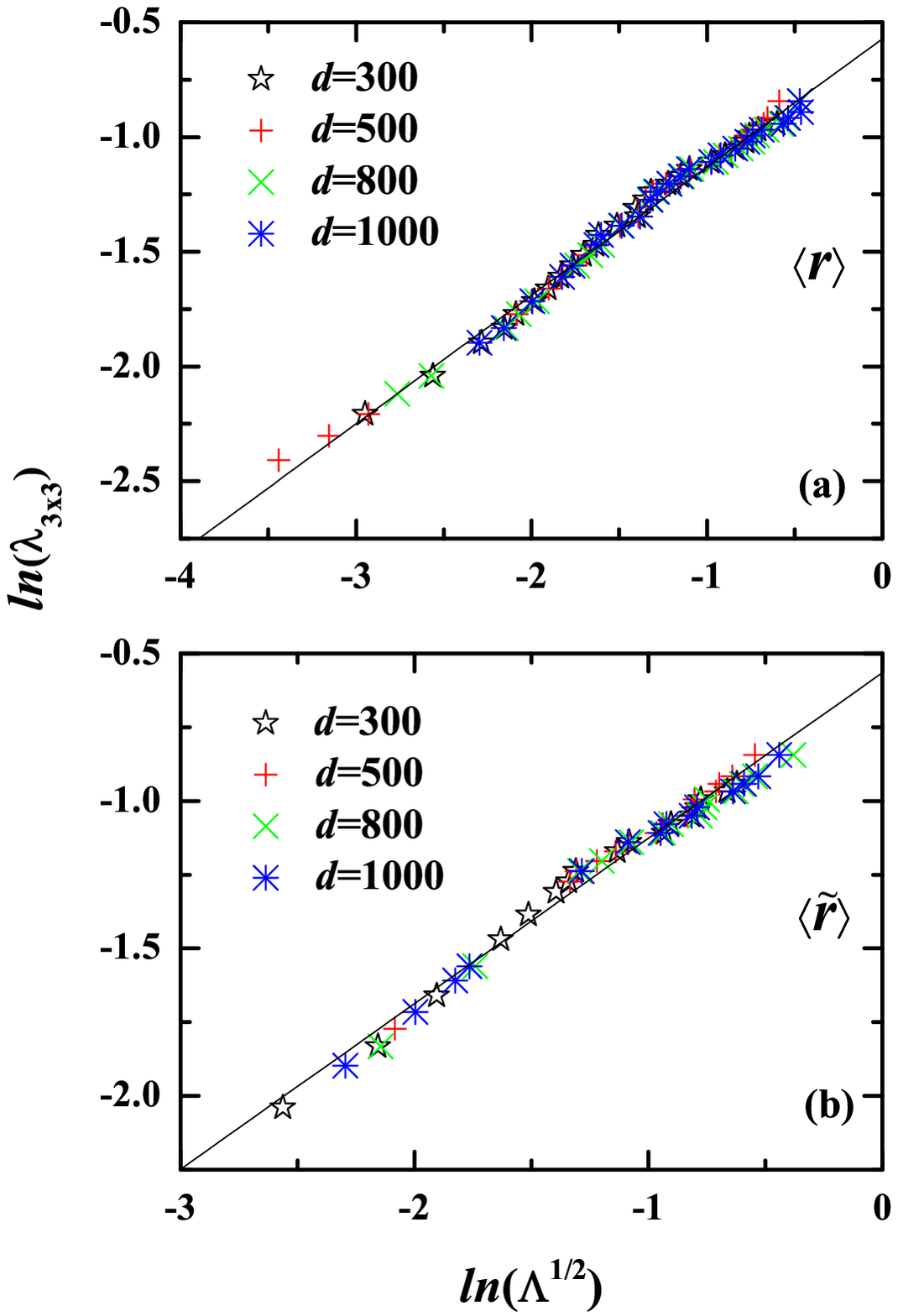}
\end{center}

\caption{(Color online) $\lambda_{3 \times 3}$ from $3 \times 3$ ensemble vs $\Lambda$ from
$d=300$ to $1000$. Upper panel gives the result obtained from  $\lan r \ran$ and
the lower panel from  $\lan \tilde{r} \ran$. The straight lines are obtained via fitting linear relation to numerical results. See text for details.}
\label{fig3}
\end{figure}

\section{A general $d \times d$ ensemble for Poisson to GOE transition:
Numerical results}
\label{sec:3}

Following the work in \cite{Karol-91}, we will consider the interpolating
Hamiltonian ensemble
\be
H_\lambda = \dis\frac{H_0 + \lambda\, V}{\dis\sqrt{1+\lambda^2}}
\label{eq6}
\ee
where $H_0$ is a diagonal matrix with $(H_0)_{ii}$, $i=1,2\ldots,d$ being
independent G(0,1) variables. Similarly $V$ is chosen as a GOE with matrix
elements variance $v^2$ (for diagonal matrix elements it is $2v^2$). Note that
the spectral variance generated by $H_0$ is $\overline{\sigma^2(H_0)} =1$ and
the one generated by $V$ is $\overline{\sigma^2(V)}=v^2(d+1)$. We choose $H_0$
and $V$ such that $\overline{\sigma^2(H_0)}= \overline{\sigma^2(V)}$ giving $v^2
\sim 1/d$ for large $d$. Similarly, as $\sigma(H_0)=1$ we have $D_0 \sim
{(d\rho_0)}^{-1}$. Here, $\rho_0$ is the eigenvalue density generated by $H_0$.
With this, for the $H$ ensemble given by Eq.~(\ref{eq6})
the transition parameter is $\Lambda = \dis\frac{\lambda^2v^2}{\l[\;\overline{
D_0}\;\r]^2} \sim \lambda^2\,d \rho_0^2$. Replacing $\rho_0$ by its average
value we have
\be
\Lambda =\lambda^2 d/2\pi
\label{eq7}
\ee
as the transition parameter. We will see ahead the significance of the
$\Lambda$ parameter in Eq.~(\ref{eq7}). As we do not have an analytical solution
for $P(r)$ for the $H_\lambda$ ensemble, we have undertaken large scale
numerical investigations with matrix dimensions changing from 300 to 1000. Instead
of studying the variation of the $P(r)$ as a function of $\lambda$ for various
values of $d$, we have considered $\lan r\ran$ and $\lan \tilde{r}\ran$ vs
$\lambda$. It is clearly seen from the results that the values of $\lan r\ran$ and $\lan \tilde{r}\ran$
depend on the matrix dimension $d$ and the transition from Poisson to GOE is faster as the value of $d$ increases. Thus,
the curves are not universal when $\lan r\ran$ and $\lan \tilde{r}\ran$ plotted against the $\lambda$ parameter. For
$d$ going to infinity, one might expect that an infinitesimal $\lambda$ would take the system to the chaotic
domain. This sudden transition with $\lambda > 0$ is similar to the situation with Poisson to GOE or GUE \cite{KS-99,FGM-1998} and
GOE to GUE \cite{PM-1983} transitions. Also, the $d$ for physical systems is ill defined. Hence, the solution
is to derive the $P(r)$ as a function of $(\lambda, d)$ and then one can identify (as in \cite{KS-99,FGM-1998,PM-1983}) the
appropriate transition parameters which will be a function of $\lambda$ and $d$. Using the transition parameter $\Lambda$ given by Eq.~(\ref{eq7}), the variation of $\lan r\ran$ and $\lan \tilde{r}\ran$ as a
function of $\Lambda$ for different values of $d$ are shown in Fig.~\ref{fig2}.
From these results, it can be seen that the variation of $\lan r\ran$ and $\lan \tilde{r}\ran$ with $\Lambda$ is slow. Similar behavior is observed in \cite{KS-99}. The continuous curves in Figs.~\ref{fig2}a and \ref{fig2}b are due to fitting exponential decay and exponential growth functions of order 3 with the numerical results respectively. It is clear from the results that
the curves obtained here are independent of the matrix dimension $d$ and thus they are universal curves.

It is possible to define a critical value $\Lambda_c$ for the transition
parameter $\Lambda$ for Poisson to GOE transition that defines the onset of
GOE fluctuations. As seen from Fig.~\ref{fig2}, the transition curves do not display phase
transition as a function of $\Lambda$. Therefore, we define the critical value of the transition parameter using a meaningful criterion. As
$\lan \tilde{r}\ran$ changes from 0.386 to 0.536 with level
fluctuations changing from Poisson to GOE, we will fix the critical value $\Lambda_c$ at $80\%$ change
in the $\lan \tilde{r}\ran$ value, which is $\lan \tilde{r}\ran=0.5$. It can be seen from Fig.~\ref{fig2} that
with this choice the critical value of $\Lambda$ is found to be $\Lambda_c=0.3$ and
this gives $\lan r\ran$=2 at $\Lambda=\Lambda_c$. This criterion can
be used to determine the onset of chaos (GOE fluctuations) in complex many-body
systems. Some examples will be discussed in Section~\ref{sec:4}.

Another important result is that, there is a mapping between the $3 \times 3$
ensemble defined by Eq.~(\ref{eq3}) and the $d \times d$ ensembles defined by Eq.~(\ref{eq6}).
 This is as follows. In order to avoid confusion between the $\lambda$ used
in Eq.~(\ref{eq3}) and Eq.~(\ref{eq6}), let us denote the $\lambda$ in $H_{3 \times
3}$ as $\lambda_{3 \times 3}$ from now onwards. Numerically, using 50000
members, the $3 \times 3$ ensemble defined by Eq.~(\ref{eq3}) was constructed,
and calculated are $\lan r\ran$ and $\lan \tilde{r}\ran$ as a function of $
\lambda_{3 \times 3}$. Using these and the results in Fig.~\ref{fig2}, one can
determine given a value of $\lambda_{3 \times 3}$, the value of $\Lambda$
(for any $d$) that gives the same value for $\lan r\ran$ and similarly
$\lan \tilde{r}\ran$. Plot of $\lambda_{3 \times 3}$ vs $\Lambda$
thus determined using $\lan r\ran$ is given in Fig.~\ref{fig3}a. Similar plot obtained
using $\lan \tilde{r}\ran$ is given in Fig.~\ref{fig3}b. It is clearly seen from Figs.~\ref{fig3}a
and 3b that the curves $\lambda_{3 \times 3}$ and $\Lambda$ are linearly
correlated.  The straight lines, in Figs.~\ref{fig3}a and~\ref{fig3}b
are due to a linear fit to $\lan r\ran$ vs $\Lambda$ and $\lan \tilde{r}\ran$ vs
$\Lambda$ curves respectively. It is interesting to note that both the results give similar slope. This indicates that
the mapping between $3 \times 3$ and $d \times d$ ensembles is same for both $\lan r\ran$ and $\lan \tilde{r}\ran$ values. Therefore the $\lan r\ran$ vs $\Lambda$ and $\lan \tilde{r}\ran$ vs $\Lambda$ transition curves can be derived by solving the $3 \times 3$ ensemble
defined by Eq.~(\ref{eq3}) and this will be addressed in future.

\section{Applications to a disordered Interacting spin-1/2 chain}
\label{sec:4}

Chains of interacting spin-1/2 systems are
prototype quantum many-body systems. They have been considered as models for
quantum computers, magnetic compounds and have recently been simulated in
optical lattices \cite{BDZ2008, Simon2011, Trotzky2012}. In the past, the
transition from integrability to chaos was investigated for a Heisenberg
spin-1/2 chain with defects using NNSD \cite{Santos2004}. Here we will consider this
model to display order to chaos transition using $P(r)$ distribution by varying
the defect strength. The Hamiltonian describing the system of a spin-1/2 chain is
\be
H =  \dis\sum_{n=1}^{L} \w_n S_n^z + \e S_d^z \; + \dis\sum_{n=1}^{L-1} J \;
\hat{S}_n \cdot \hat{S}_{n+1}.
\label{eq8}
\ee
Here $L$ is the number of sites and $\hbar=1$. $\hat{S}_n = \vec{\sigma}_n/2$
are the spin operators at site $n$, and $\vec{\sigma}$ are the Pauli spin
matrices. The first term in Eq.~(\ref{eq8}) corresponds to the Zeeman splitting of each spin
$n$ determined by the static magnetic field in the $z$-direction giving the
energy splitting $\w_n$. If all the sites are assumed to have same energy splitting
$\w$ except a single site $d$, where the splitting is $\w+\varepsilon$ then this
site is referred to as the defect (second term in Eq.~(\ref{eq8})). If $\w_n=\w$ for all
sites with $\varepsilon=0$  then the chain is called clean. The last term in Eq.~(\ref{eq8})
describes two types of coupling between the nearest neighbor spins. First is the
diagonal Ising interaction $(S_n^z S_{n+1}^z)$ and second is the off-diagonal
flip-flop term $(S_n^x S_{n+1}^x + S_n^y S_{n+1}^y)$. The latter term is
responsible for propagating the excitation through the chain.
To avoid degeneracy found in the spectrum of a close chain (chain with periodic boundaries), in the present analysis
we have used an open chain (with free boundaries). Further more, we have considered isotropic chain, i.e. coupling strength between the
Ising interaction is equal to that of the flip-flop term and it is taken as
constant. These 1D systems of interacting spin-1/2 chain are devoid
of random elements and involve only two-body interactions.

For the chain described by Eq.~(\ref{eq8}), the $z$ component of the total spin
$\sum_{n=1}^L S_n^z$ is conserved, so the states with same number of excitations are
coupled. Here the focus is on $L/2$ excitations giving largest number of states.
In the absence of defects (when the chain is clean), spin-1/2 chain is
integrable and is solved with the Bethe ansatz \cite{Bethe}.  An open chain with
defects only on the edges is also integrable \cite{Alcaraz}. The NNSD for such an
interacting chain follows the Poisson distribution. On the other hand, when there is
only one defect in the middle of the chain and the defect excess energy is of
the order of the interaction strengths, the NNSD for such a chain is described by the
Wigner (GOE) distribution \cite{Santos2004}. When the defect strength is very strong
compared to that of the interaction term, then an excitation on one side of
the chain would not have enough energy to overcome the defect and reach the
other side of the chain. In this situation, the chain is said to be broken giving
an integrable model and Poisson form for the NNSD is recovered \cite{Santos2004}.
Here, to display the Poisson to GOE transition, the evolution of $P(r)$ is
studied with respect to the defect strength with $\varepsilon \leq J$ and taking defect site at the middle of
the chain with $d=L/2$.

\begin{figure}[tbh]
\begin{center}
\includegraphics[width=0.6\linewidth]{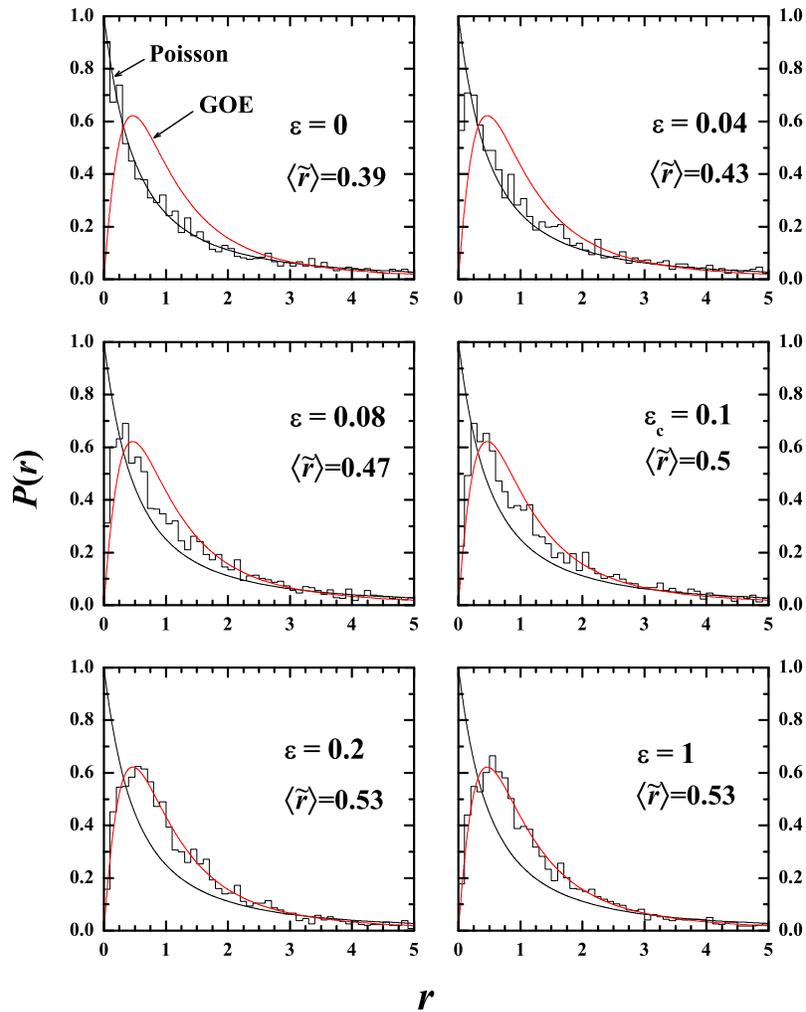}
\end{center}

\caption{(Color online) Histograms represent $P(r)$ distribution, for $H$ in Eq.~(\ref{eq8}) with $L=14$
with 7 excitations, for various values of defect strength $\varepsilon$. The
defect is on the site $d=7$. In applying Eq.~(\ref{eq8}), the value of $\omega_n$ is
chosen to be $\omega_n=0$ for all $n$ and similarly, $J$ value is chosen to be
$J=1$. Values of $\lan \tilde{r}\ran$ are also given in the figure and $\lan
\tilde{r}\ran=0.5$ gives value of chaos marker $\varepsilon_c$. Bin size equal to
0.1 is used for the histograms. The results are compared with Poisson and
Wigner(GOE) predictions.}
\label{fig4}
\end{figure}

\begin{figure}[tbh]
\begin{center}
\includegraphics[width=0.4\linewidth]{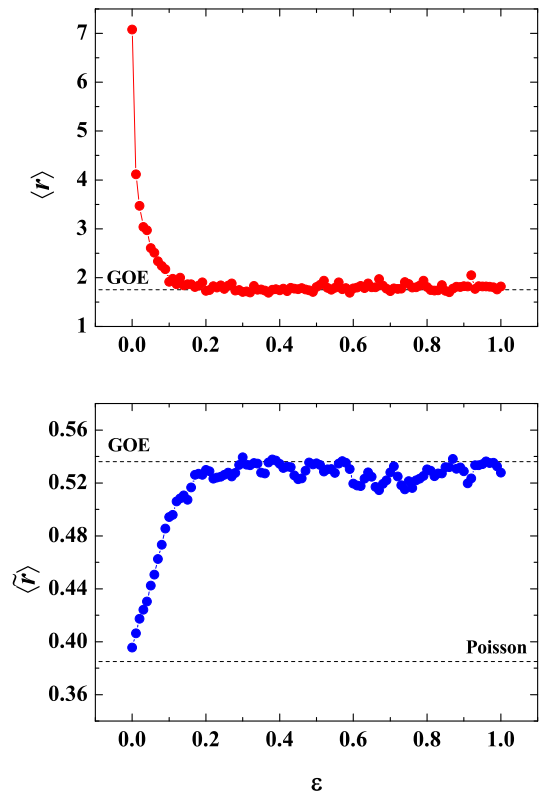}
\end{center}

\caption{(color online) Average value of $r_n$ (denoted as $\lan r\ran$)(upper panel)  and
$\tilde{r}_n$ (denoted as $\lan \tilde{r}\ran$)(lower panel) as a function of
defect strength $\varepsilon$ for $H$ in Eq.~(\ref{eq8}) with $L=14$, 7 spins up and
defect site $d=7$.}
\label{fig5}
\end{figure}

Fig.~\ref{fig4} represents $P(r)$ histograms constructed, using $L=14$ sites and $7$
excitations, for various values of $\varepsilon$. The dimensionality of the
system is $3432$. The figure clearly displays the transition in $P(r)$ from the
Poisson character for low values of $\varepsilon$ to the GOE character as the
strength of the defect is slowly increased by increasing $\varepsilon$. The
transition of $P(r)$ is smooth in character with respect to the
parameter $\varepsilon$. Fig.~\ref{fig5} shows variation of $\lan r\ran$ and $\lan
\tilde{r}\ran$ as a function of the parameter $\varepsilon$. Demarcation between
the Poisson to GOE region giving the critical value of the defect strength, at which there is a clear onset of
the transition to GOE,  is found to be $\varepsilon_c=\varepsilon(\lan\tilde{r}\ran=0.5)=0.1$ for the example
considered here. The results in Fig.~\ref{fig5} are similar to
those in Fig.~\ref{fig2} confirming that the random
matrix model considered in Section~\ref{sec:3} will be appropriate for the 1D spin-1/2
chain system with $H$ defined by Eq.~(\ref{eq8}).

\section{Conclusions}
\label{sec:5}

Going beyond the results in \cite{CK}, in the present work we have studied the Poisson to GOE
transition in the distribution $P(r)$ of the ratio of consecutive level
spacings and the related averages $\lan r\ran$ and  $\lan \tilde{r}\ran$. Main
conclusions are: (i) there is a scaling in $P(r)$, $\lan r\ran$ and  $\lan
\tilde{r}\ran$ measures similar to the scaling seen before for NNSD and  related
averages such as the variance of NNSD giving a transition parameter $\Lambda$
defined by Eq.~(\ref{eq7}); (ii) universal transition curves for  $\lan r\ran$ and
$\lan \tilde{r}\ran$ vs $\Lambda$ are constructed numerically as given in Fig.~\ref{fig2};
 (iii) as seen from Fig.~\ref{fig3}, the $3 \times 3$ random matrix ensemble given by
Eq.~(\ref{eq3}) maps to the $d \times d$
matrices for $P(r)$ and related averages. This is similar to the mapping to a $2 \times 2$ matrix for NNSD \cite{Ha-91,Karol-91,KS-99}.
As an example for Poisson to GOE transition in $P(r)$ and the related averages,
results for a 1D interacting spin-1/2 chain are presented.  In
future, it is important to solve analytically the $3 \times 3$ matrix ensemble
defined by Eq.~(\ref{eq3}). Finally, the results presented in this paper should be useful in applications for systems discussed in \cite{Huse2007,OPH2009,Pal-10,Iyer-12,koll2010,Coll2012,Torr2014}.

\section*{Acknowledgments}
We thank Soumik Bandyopadhyay for collaboration in the initial stages of this
work. NDC acknowledges support from University Grant Commission (UGC),
India [grant No: F.40-425/2011(SR)]. HND acknowledges support from UGC, India
for a research fellowship [grant No. F. 7-198/2007(BSR)].

\ed
\begin{thebibliography}{00}

\bibitem{Haake} {F. Haake, Quantum Signatures of Chaos, 3'rd
edition, Springer-Verlag, Heidelberg, 2010.}

\bibitem{stockmann} {H--J. St\"{o}ckmann, Quantum Chaos: An Introduction, 1'st
edition, Cambridge University Press, Cambridge, 1999.}

\bibitem{wimberger} {S. Wimberger, Nonlinear Dynamics and Quantum Chaos: An Introduction, Springer International Publishing, Switzerland, 2014.}

\bibitem{Berry}  {M. V. Berry, M. Tabor, Proc. Roy. Soc. (London)  A356
(1977) 375.}

\bibitem{Bohigas} {O. Bohigas, M.-J. Giannoni, C. Schmit, Phys. Rev.
Lett. 52 (1984) 1.}

\bibitem{Huse2007} {V. Oganesyan, D. A. Huse, Phys. Rev. B 75 (2007) 155111.}

\bibitem{OPH2009} {V. Oganesyan, A. Pal, D. A. Huse, Phys. Rev. B 80 (2009) 115104.}

\bibitem{Pal-10} {A. Pal, D. A. Huse, Phys. Rev. B 82 (2010) 174411.}

\bibitem{Iyer-12} {S. Iyer, V. Oganesyan, G. Refael, D. A. Huse, Phys. Rev.
B 87 (2013) 134202.}

\bibitem{koll2010} {C. Kollath, G. Roux, G. Biroli, A. M. L\"{a}uchli, J.
Stat. Mech. (2010) P08011.}

\bibitem{Coll2012} {M. Collura, H. Aufderheide, G. Roux, D. Karevski, Phys.
Rev. A 86 (2012) 013615.}

\bibitem{CK} {N.D. Chavda, V.K.B. Kota, Phys. Lett. A 377 (2013) 3009.}

\bibitem{HBCK} {S.K. Haldar, B. Chakrabarti, N.D. Chavda, T.K. Das, S. Canuto, V. K. B. Kota, Phys. Rev. A 89 (2014) 043607.}

\bibitem{RMJ} {A. Rai, V. Menon, S. Jalan, private communication.}

\bibitem{rmt2} {M.L. Mehta, Random Matrices, 3rd Edition, Elsevier B.V., Netherlands, 2004.}

\bibitem{rmt3} {G. Akemann, J. Baik, P. Di Francesco (Editors), The Oxford Handbook of Random Matrix Theory, Oxford University Press, Oxford,
2011.}

\bibitem{vkbk01} {V. K. B. Kota, Phys. Rep. 347 (2001) 223.}

\bibitem{Br-72} {T.A. Brody, Lett. Nuovo. Cim. 7 (1973) 482.}

\bibitem{Ko-14} {V.K.B. Kota, Embedded Random Matrix Ensembles in Quantum
Physics, Lecture Notes in Physics, volume 884, Springer, Heidelberg, 2014.}

\bibitem{ABGR-2013} {Y. Y. Atas, E. Bogomolny, O. Giraud, G. Roux, Phys.
Rev. Lett. 110 (2013) 084101.}

\bibitem{Atas13} {Y. Y. Atas, E. Bogomolny, O. Giraud, P. Vivo, E. Vivo, J. Phys. A: Math.
Theor. 46 (2013) 355204.}

\bibitem{Ha-91} {G. Lenz, F. Haake, Phys. Rev. Lett. 65 (1991) 2325.}

\bibitem{Karol-91} {G. Lenz, K. Zyczkowski, D. Saher, Phys. Rev. A 44,
(1991) 8043.}

\bibitem{KS-99} {V.K.B. Kota, S. Sumedha, Phys. Rev. E 60 (1999) 3405.}

\bibitem{FGM-1998} {K.M. Frahm, T. Guhr, A. M\"{u}ller-Groeling, Ann. Phys. (N.Y.) 270 (1998) 292.}

\bibitem{PM-1983} {A. Pandey, M. L. Mehta, Commun. Math. Phys. 87 (1983) 449.}

\bibitem{BDZ2008}{I. Bloch, J. Dalibard, W. Zwerger, Rev. Mod. Phys. 80 (2008) 885.}

\bibitem{Simon2011} {J. Simon, W. S. Bakr, R. Ma, M. E. Tai, P. M. Preiss, M. Greiner, Nature 472 (2011) 307.}

\bibitem{Trotzky2012} {S. Trotzky, Yu-Ao Chen, A. Flesch, Ian P. McCulloch, U. Schollwöck, J. Eisert,
I. Bloch, Nature Physics 8 (2012) 325.}

\bibitem{Santos2004} {L. F. Santos, J. Phys. A 37 (2004) 4723.}

\bibitem{Bethe} {H. A. Bethe, Z. Phys. 71 (1931) 205.}

\bibitem{Alcaraz} {F. C. Alcaraz, M. N. Barber, M. T. Batchelor, R. J. Baxter,
G. R. W. Quispel, J. Phys. A 20 (1987) 6397.}

\bibitem{Torr2014} {E.J. Torres-Herrera, L.F. Santos, Phys. Rev. A 89 (2014) 043620.}

\end{thebibliography}
